\documentclass[reprint, superscriptaddress, secnumarabic, amssymb, nobibnotes, aps, prl]{revtex4-1}

\usepackage{graphicx}
\usepackage{epstopdf}
\usepackage[T1]{fontenc}
\usepackage[latin9]{inputenc}
\usepackage{amsbsy}
\usepackage{gensymb}
\usepackage[T1]{fontenc}
\usepackage[latin9]{inputenc}
\usepackage{amsmath}
\usepackage{amssymb}
\usepackage{bbm}
\usepackage{braket}
\usepackage{xcolor}
\allowdisplaybreaks
\usepackage{graphicx}
\usepackage[colorlinks=true]{hyperref}  
\usepackage[normalem]{ulem}

\newcommand{\figref}[1]{Fig.~\ref{#1}}

\renewcommand{\approx}{\simeq}

\begin{document}
\title{\textrm{Type-I superconductivity in noncentrosymmetric superconductor AuBe}}
\author{D. Singh}
\affiliation{Indian Institute of Science Education and Research Bhopal, Bhopal, 462066, India}
\affiliation{ISIS facility, STFC Rutherford Appleton Laboratory, Harwell Science and Innovation Campus, Oxfordshire, OX11 0QX, UK}
\author{A. D. Hillier}
\affiliation{ISIS facility, STFC Rutherford Appleton Laboratory, Harwell Science and Innovation Campus, Oxfordshire, OX11 0QX, UK}
\author{R. P. Singh}
\email[]{rpsingh@iiserb.ac.in}
\affiliation{Indian Institute of Science Education and Research Bhopal, Bhopal, 462066, India}

\date{\today}
\begin{abstract}
\begin{flushleft}

\end{flushleft}
The noncentrosymmetric superconductor AuBe have been investigated using the magnetization, resistivity, specific heat, and muon-spin relaxation/rotation measurements. AuBe crystallizes in the cubic FeSi-type B20 structure with superconducting transition temperature observed at $T_{c}$ = 3.2 $\pm$ 0.1 K. The low-temperature specific heat data, $C_{el}$(T), indicate a weakly-coupled fully gapped BCS superconductivity with an isotropic energy gap 2$\Delta(0)/k_{B}T_{c}$ = 3.76, which is close to the BCS value of 3.52. Interestingly, type-I superconductivity is inferred from the $\mu$SR measurements, which is in contrast with the earlier reports of type-II superconductivity in AuBe. The Ginzburg-Landau parameter is $\kappa_{GL}$ = 0.4 $<$ 1/$\sqrt{2}$. The transverse-field $\mu$SR data transformed in the maximum entropy spectra depicting the internal magnetic field probability distribution, P(H), also confirms the absence of the mixed state in AuBe. The thermodynamic critical field, $H_{c}$, calculated to be around 259 Oe. The zero-field $\mu$SR results indicate that time-reversal symmetry is preserved and supports a spin-singlet pairing in the superconducting ground state.

\end{abstract}
\maketitle

\section{INTRODUCTION}

Noncentrosymmetric (NCS) superconductors (SCs) which lack inversion symmetry in the crystal structure display a variety of unusual properties in the superconducting state \cite{BOOK}. In superconductors with conserved inversion symmetry, the Cooper-pair wave function is strictly determined by the parity symmetry. This means that the pair function generally consists of either spin-singlet or spin-triplet as the spin part and  s, p, or d,...-wave as the the orbital part \cite{PWA}. However, for NCS SCs the conventional Cooper pairs can no longer form. The lack of inversion symmetry in NCS SCs gives rise to asymmetric spin-orbit coupling (ASOC) which causes the splitting of Fermi surface \cite{BOOK}. This splitting allows for mixed spin-singlet and spin-triplet pairing \cite{BOOK,VME,VME1,LPG}. Mixed-parity superconductivity can lead to various exotic unconventional superconducting properties in  noncentrosymmetric materials \cite{VME,VME1,LPG,SKYP,SFJ}. \\

Evidence of unconventional superconducting properties has been observed in several NCS SCs, for example: CePt$_{3}$Si \cite{IBW}, Li$_{2}$Pt$_{3}$B \cite{HQY,MNY} and CeIrSi$_{3}$ \cite{HMT} exhibit lines nodes in the superconducting gap, whereas LaNiC$_{2}$ \cite{JCL} and (La,Y)$_{2}$C$_{3}$ \cite{SKY} show multigap superconductivity. Unusually high upper critical field compared to the Pauli paramagnetic limiting field was observed in CePt$_{3}$Si\cite{EBG} and Ce(Rh,Ir)Si$_{3}$ \cite{NKK,ISY}, due to the influence of strong ASOC. Furthermore, $\mu$SR measurements have found time-reversal symmetry (TRS) breaking in LaNiC$_{2}$ \cite{ADH}, Re$_{6}$(Ti,Zr,Hf) \cite{DSJ1,RPS,DSJ}, La$_{7}$Ir$_{3}$ \cite{JAT} and SrPtAs \cite{PKB1}. However, other systems such as: BiPd \cite{KMS,XBY}, Nb$_{0.18}$Re$_{0.82}$ \cite{ABK}, LaMSi$_{3}$ (M = Rh, Ir)\cite{VKA1,VKA2}, LaMSi$_{3}$ (M = Pd, Pt)\cite{MSA3}, T$_{2}$Ga$_{9}$ (T = Rh, Ir) \cite{KWS,TSM} appear to behave as conventional s-wave fully-gapped superconductors. These varied properties make it valuable to study additional noncentrosymmetric systems in an effort to gain a deeper understanding of their physics.\\

Recently, the physical properties of NCS superconductor AuBe with superconducting transition temperature $T_{c}$ of 3.3 K reported by Amon \textit{et al.} \cite{AuBe}. The discovery of superconductivity in AuBe was first reported by Matthias \textit{et al.} in early 1960 \cite{BTM}. However, no comment was made regarding the nature of superconductivity and noncentrosymmetry in this material. AuBe has structural phase transition at $T_{s}$ = 80 K, where it undergoes transition from the high-temperature CsCl structure to the low-temperature FeSi structure. As represented in Ref. \cite{AuBe} that in the low-temperature phase (FeSi structure) the Au atoms are surrounded by 7 Be atoms and 6 Au atoms whereas in high-temperature phase (CsCl structure) Au atoms are coordinated by 8 Be atoms and 6 Au atoms.  It is the cubic FeSi-type B20 structure which is noncentrosymmetric and thoroughly investigated in this work.  Interestingly, the superconducting properties of AuBe investigated by transport, magnetic, specific heat measurements suggests a weakly coupled type-II superconductivity with lower and upper critical field 32 Oe and 335 Oe respectively \cite{AuBe}. However, it is noted that the physical properties of AuBe are often challenging to entirely define the class of superconductivity. Therefore, it warrants an in-depth analysis of the superconducting state of AuBe using the muon spin rotation/relaxation ($\mu$SR) measurements. This method has been proven to be successful in determining the probability distribution of the internal fields which subsequently can provide various types of information on the SC state.\\ 

In this paper, we have investigated the superconducting properties of AuBe using resistivity, magnetization, specific heat and $\mu$SR measurements. The transverse-field (TF)-$\mu$SR asymmetry spectra were transformed into a probability of field versus magnetic field diagram by a Maximum Entropy (MaxEnt) algorithm \cite{MAX} to observe the dominant field components. Using this information, the TF-$\mu$SR spectra in the time domain were analysed using a sum of Gaussian field distributions to quantify the different magnetic field distribution present in the sample as a function of temperature and field. In addition, the zero-field (ZF)-$\mu$SR measurements were employed to search for TRS breaking phenomena in the material. Interestingly, our results suggest that AuBe can be classified as a weakly-coupled type-I superconductor with a thermodynamic critical field $H_{c}$ $\approx$ 259 $\pm$ 1 Oe in contrast with the earlier reports \cite{AuBe}. The Ginzburg -Landau parameter $\kappa_{GL}$ obtained was around 0.4 $<$ 1/$\sqrt{2}$ again confirming type-I superconductivity in the system.

\section{EXPERIMENTAL METHODS}
The polycrystalline samples of AuBe were prepared by the standard arc melting of stoichiometric quantities of the elements Au (5N, Alfa Aesar) and Be (5N, Alfa Aesar) on a water-cooled copper hearth under a high purity argon gas atmosphere.  Arc melting was done 2-3 times sequentially without removal from the argon gas atmosphere. The mass loss ($<$ 2$\%$) is checked after the melting and a small amount of beryllium was added to compensate for the Be loss before the final melting. The as-cast samples were then sealed in an evacuated quartz tube and annealed at 500 $^{\circ}$C for 48 h.\\
Room temperature powder x-ray diffraction measurements were carried out on a PANalytical diffractometer using Cu K$_{\alpha}$ radiation ($\Lambda$ = 1.54056 $\text{\AA}$). Temperature and field dependent magnetization measurements were made using a Quantum Design superconducting quantum interference device (SQUID MPMS 3, Quantum Design). The measurements were performed in the temperature range of 1.8 K to 4.0 K with applied magnetic fields up to 300 Oe. Specific heat measurements were performed by the two tau time-relaxation method using the physical property measurement system (PPMS, Quantum Design, Inc.) in zero applied magnetic field. The electrical resistivity was measured by a conventional four-probe technique using the PPMS in a temperature range 1.8 K to 300 K and in fields up to 200 Oe.\\
The $\mu$SR measurements were carried out using the MUSR spectrometer at the ISIS Neutron and Muon facility, in STFC Rutherford Appleton Laboratory, United Kingdom. The powdered AuBe sample was mounted on a high-purity-silver plate using diluted GE varnish. The measurements were performed in the temperature range 0.1 K - 3.5 K using a sorption He3 cryostat. The $\mu$SR measurements were performed under zero-field and transverse-field conditions. A full description of the $\mu$SR technique may be found in Ref. \cite{SLL}. In ZF-$\mu$SR, the contribution from the stray fields at the sample position due to neighbouring instruments and the Earth's magnetic field is cancelled to within $\sim$ 1.0 $\mu$T using three sets of orthogonal coils. TF-$\mu$SR measurements are performed to investigate the magnetic field distribution inside the sample. In particular, the TF-$\mu$SR data were analyzed using Maximum Entropy technique to determine the probability distribution, P(H), of the internal magnetic fields. Applied fields between 50 and 300 Oe were used to fully cover the superconducting phase diagram of AuBe.

\section{RESULTS AND DISCUSSION}
\begin{figure}
\includegraphics[width=1.0\columnwidth]{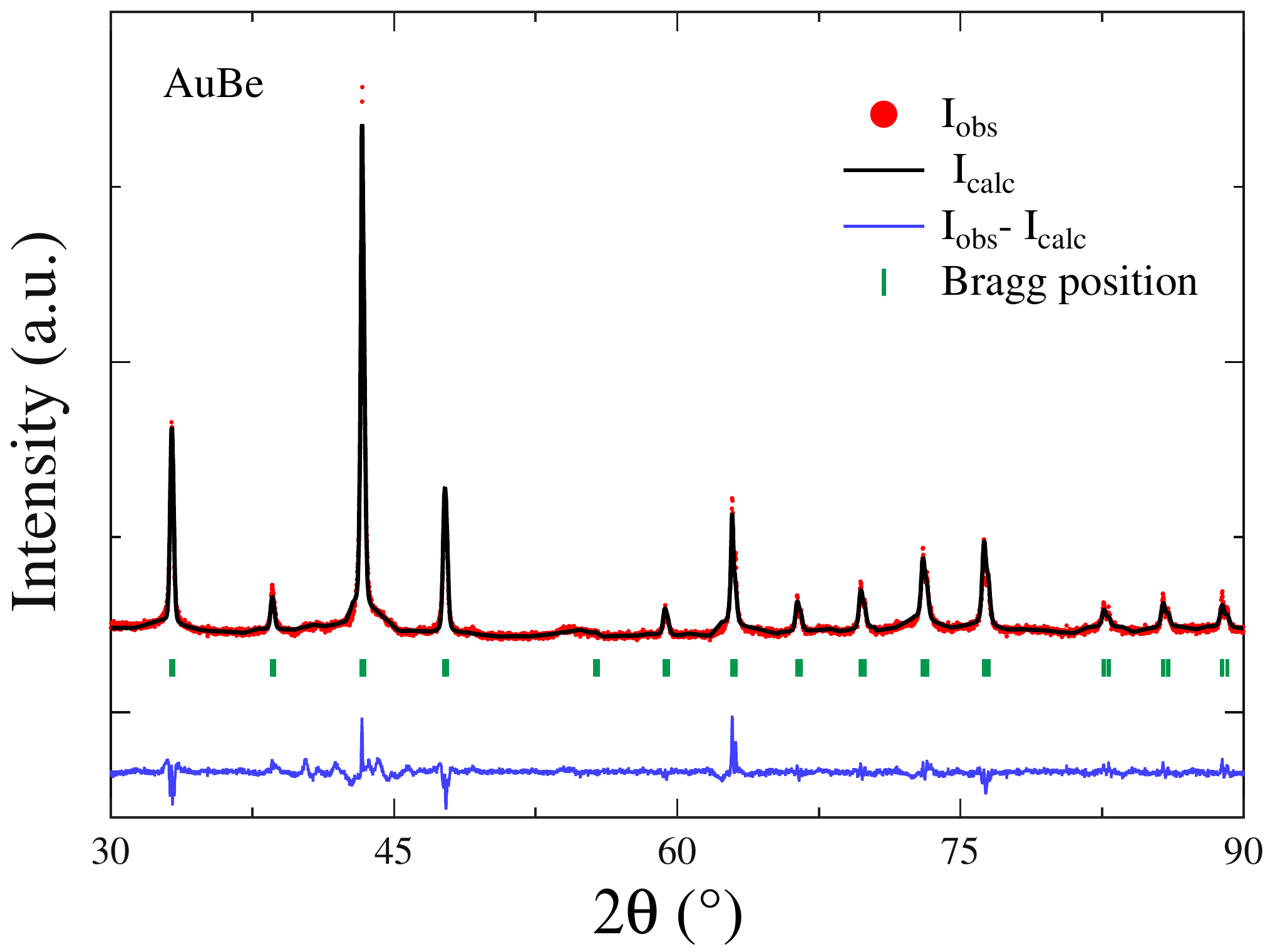}
\caption{\label{Fig1:Fig1}The powder x-ray diffraction pattern of AuBe at room temperature. The solid black line is a Rietveld refinement to the data. The vertical tick marks indicate the calculated peak positions, and the lower graph shows the difference plot.}
\end{figure}

\begin{figure}
\includegraphics[width=1.0\columnwidth]{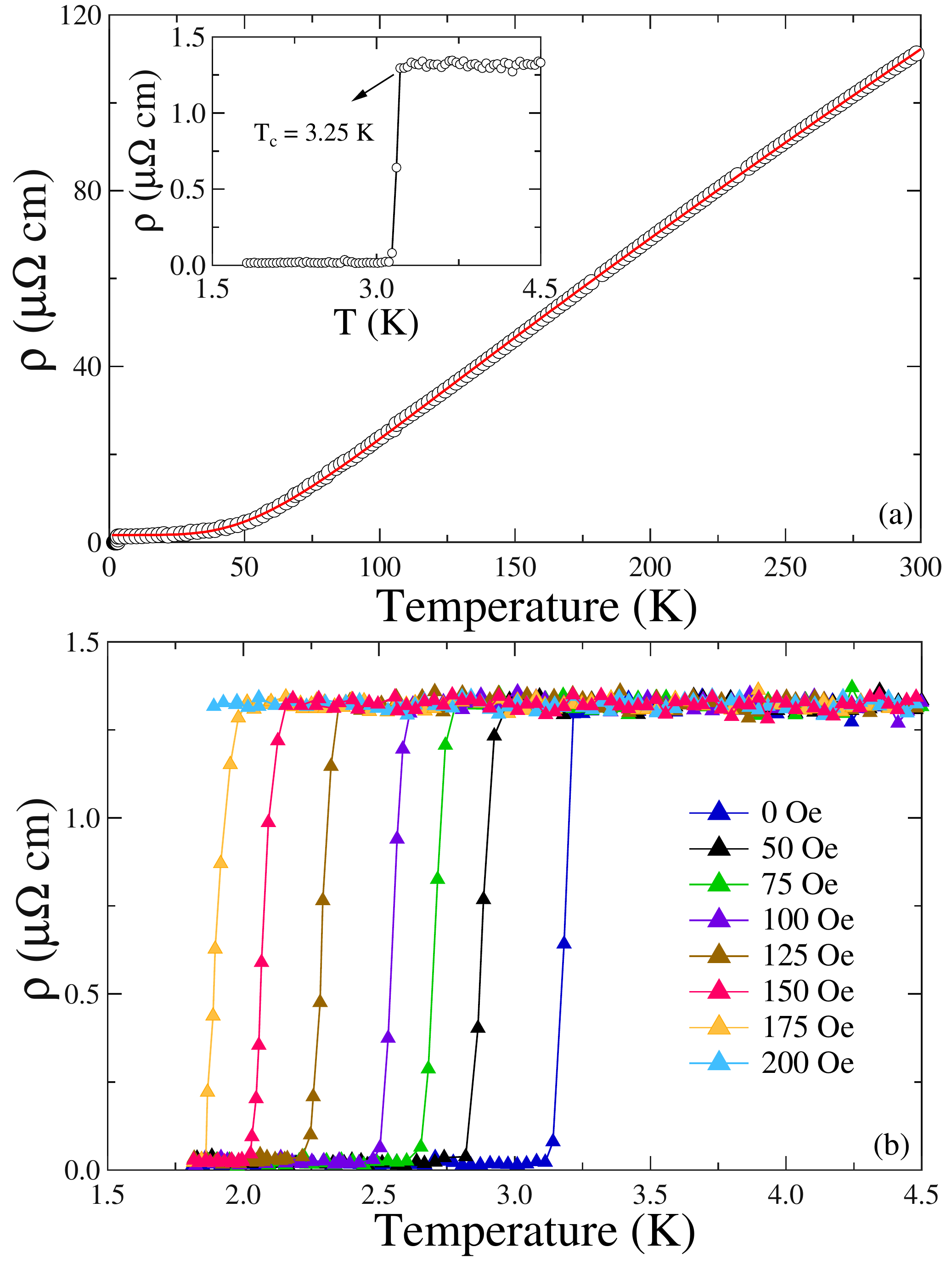}
\caption{\label{Fig2:Fig2}(a) Temperature dependence of electrical resistivity $\rho(T)$ of AuBe for 1.8 K $\le$ T $\le$ 300 K measured in zero applied magnetic field. The red solid curve is a fit of $\rho(T)$ data by the Bloch-Gr$\ddot{\mathrm{u}}$neisen model. The inset shows the expanded view of $\rho(T)$ with superconductivity at $T_{c~onset}$ = 3.25 K. (b) $\rho(T)$ of AuBe for
1.8 K $\le$ T $\le$ 4.5 K, showing the superconducting transitions for different values of H.}
\end{figure}

Figure \ref{Fig1:Fig1} shows the room temperature powder x-ray diffraction pattern of the synthesized sample of AuBe. The refinement shows that our sample crystallizes into a single phase of the expected noncentrosymmetric cubic FeSi-type structure with the unit cell parameter a = 4.6684(3) $\text{\AA}$. Small impurity phase is observed in the XRD pattern, however, no significant effect of this impurity phase is observed in the bulk and muon spectroscopy measurements. These results are in good agreement with the published literature \cite{AuBe}.\\
Electrical resistivity $\rho$ versus temperature T data for AuBe was measured in  zero applied magnetic field within the temperature range 1.8 K $\le$ T $\le$ 300 K, as shown in \figref{Fig2:Fig2}(a). The metallic nature of the sample can be inferred from the T dependence of $\rho$, where resistivity decreases consistently with decreasing temperature. At T = 300 K, the value of resistivity is $\rho(300 K)$ $\approx$ 112 $\mu\Omega$ cm and the  resistivity at T = 4 K is $\rho(4 K)$ $\approx$ 1.35 $\mu\Omega$ cm, giving a residual resistivity ratio (RRR) $\approx$ 83. The low value of residual resistivity just above the superconducting state along with the high value of RRR reflects the good quality of our sample. The inset of \figref{Fig2:Fig2}(a) highlights the sharp drop to zero resistance below $T_{c~onset}$ = 3.25 K, signalling the onset of superconductivity in AuBe. The zero resistivity value is reached at $T_{c~0}$ = 3.15 K. The transition temperature, $T_{c}$, which is defined as the midpoint of the transition is 3.2 $\pm$ 0.1 K, close to the value published previously \cite{AuBe}. The $\rho(T)$ measurements were performed at various applied magnetic fields between 0 $\le$ H $\le$ 200 Oe as shown in \figref{Fig2:Fig2}(b). The application of magnetic field suppresses the superconducting transition temperature T$_{c}$ rapidly; at H = 200 Oe, T$_{c}$ decreases below 1.8 K from 3.17 K at H = 0.\\ 
\begin{figure}
\includegraphics[width=1.0\columnwidth]{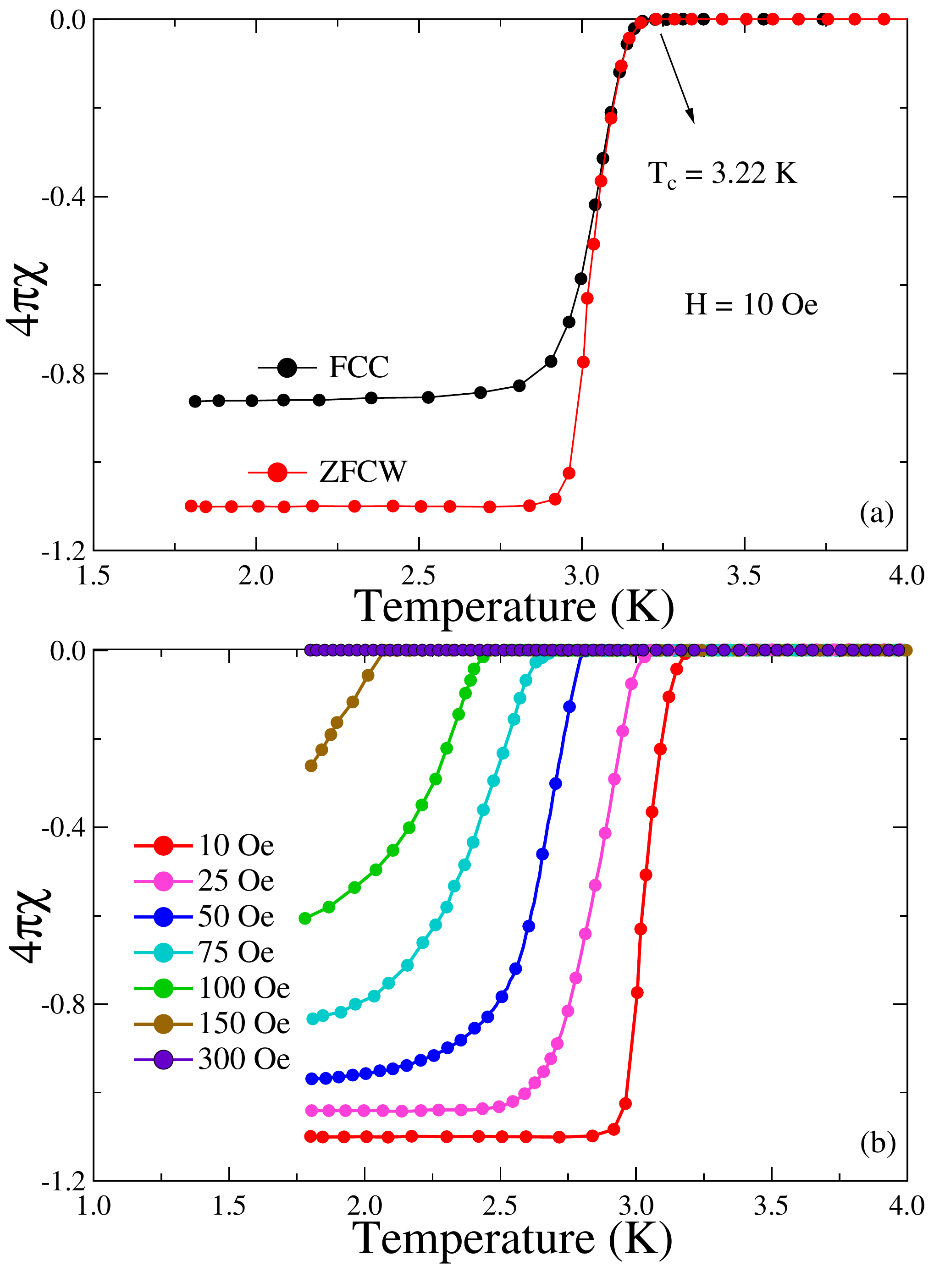}
\caption{\label{Fig3:Fig3}(a) Temperature dependence of dc susceptibility $\chi(T)$  in ZFCW and FCC mode show the superconductivity at $T_{c~onset}$ = 3.22 K. (b) The $\chi(T)$ measurements were done at different applied magnetic fields between 0 Oe $\le$ H $\le$ 300 Oe.}
\end{figure} 
The normal-state resistivity of AuBe is analyzed using the Bloch-Gr$\ddot{\mathrm{u}}$neisen (BG) model, which describes the resistivity arising due to electrons scattering from the acoustic phonons. The temperature dependence of the resistivity,  $\rho$(T), is modeled as 
\begin{equation}
 \rho(T) = \rho_{0} + \rho_{BG}(T)
\label{para2}
\end{equation}
where $\rho_{0}$ is the residual resistivity due to the defect scattering and is essentially temperature independent whereas $\rho_{BG} $ is the BG resistivity given by \cite{GG}
\begin{equation}
 \rho_{BG}(T) = 4C\left(\frac{T}{\Theta_{R}}\right)^{5}\int_{0}^{\Theta_{R}/T}\frac{x^{5}}{(e^{x}-1)(1-e^{-x})}dx  
\label{para3}
\end{equation}
where $\Theta_{R}$ is the Debye temperature obtained from resistivity measurements, while C is a material dependent pre-factor \cite{ABA}. The best fit for the above data using the BG model is shown by the solid red curve in \figref{Fig2:Fig2}(a), and yields a Debye temperature $ \Theta_{R} $ =  (345 $ \pm $ 2) K, C = (136 $ \pm $ 15) $\mu\Omega $cm and residual resistivity $\rho_{0}$ = ( 1.36 $\pm$ 0.05) $ \mu\Omega $cm. The value of the Debye temperature $\Theta_{R}$ is close to that extracted later from the specific heat measurements.\\ 
The mean free path, $l$, can be calculated using the Fermi velocity $v_{F}$ and the scattering time $\tau$ using the relation $l$ = $v_{F}$$\tau$. Based on the Drude model, $v_{F}$ is given by $\hbar$$k_{F}$/m$^{*}$ and scattering time $\tau^{-1}$ = ne$^{2}$$\rho_{0}$/m$^{*}$, where m$^{*}$ is the effective mass, $k_{F}$ is the Fermi vector and n is the electron carrier density. As already shown in Ref. \cite{AuBe}, the electron density for AuBe is n = 4/V = 3.93 $\times$ 10$^{28}$ $m^{-3}$ for V = 101. 746 \text{\AA}$^{3}$. Assuming a spherical Fermi surface, we can use the above value of n to estimate then Fermi wave vector $k_{F}$ = (3n$\pi^{2}$)$^{1/3}$ = 1.05 $\text{\AA}^{-1}$. The effective mass m$^{*}$ is estimated to be 2.8$m_{e}$ , using the values for $\gamma_{n}$ (from specific heat measurement), $k_{F}$ and n \cite{AuBe}. From the calculated values of m$^{*}$, n, $k_{F}$ and $\rho_{0}$, we determined Fermi velocity $v_{F}$ = 4.34 $\times$ 10$^{5}$ m/s and mean free path $l$ = 808.2 \text{\AA}.\\  
Superconductivity in AuBe was further confirmed by dc susceptibility ($\chi$) measurement taken in zero-field cooled warming (ZFCW) and field-cooled cooling (FCC) mode as shown in \figref{Fig3:Fig3}(a). The data were measured with an applied field of 10 Oe, demonstrating the onset of a sharp superconducting transition at T$_{c~onset}$ = 3.22 $\pm$ 0.1 K. The value of Meissner fraction 4$\pi \chi$ exceeds 100 $\%$ due to the demagnetiztion effects \cite{AA}, however, it indicates complete flux expulsion in the compound. The plot for $\chi(T)$ measurements in several applied magnetic fields up to 300 Oe is shown in \figref{Fig3:Fig3}(b). Upon the application of the magnetic field, $T_{c~onset}$ is strongly suppressed where $T_{c~onset}$ becomes smaller than 1.8 for H $>$ 150 Oe.
\begin{figure}
\includegraphics[width=1.0\columnwidth]{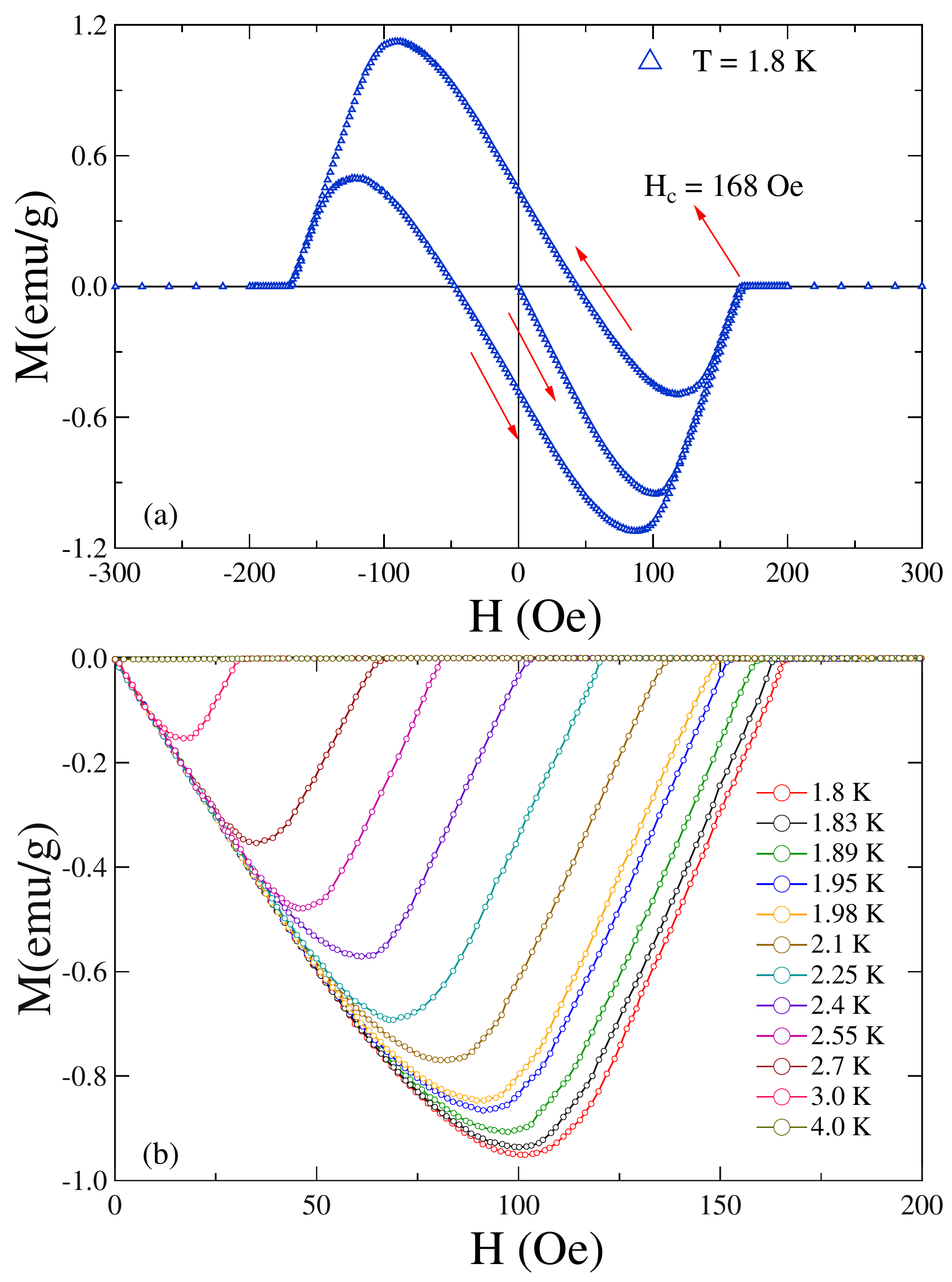}
\caption{\label{Fig4:Fig4}(a) Magnetization as a function of applied field measured at 1.8 K. (b) Magnetic isotherms, M(H), for the temperature range 1.8 K $\le$ T $\le$ 4.0 K.}
\end{figure}
\begin{figure}
\includegraphics[width=1.0\columnwidth]{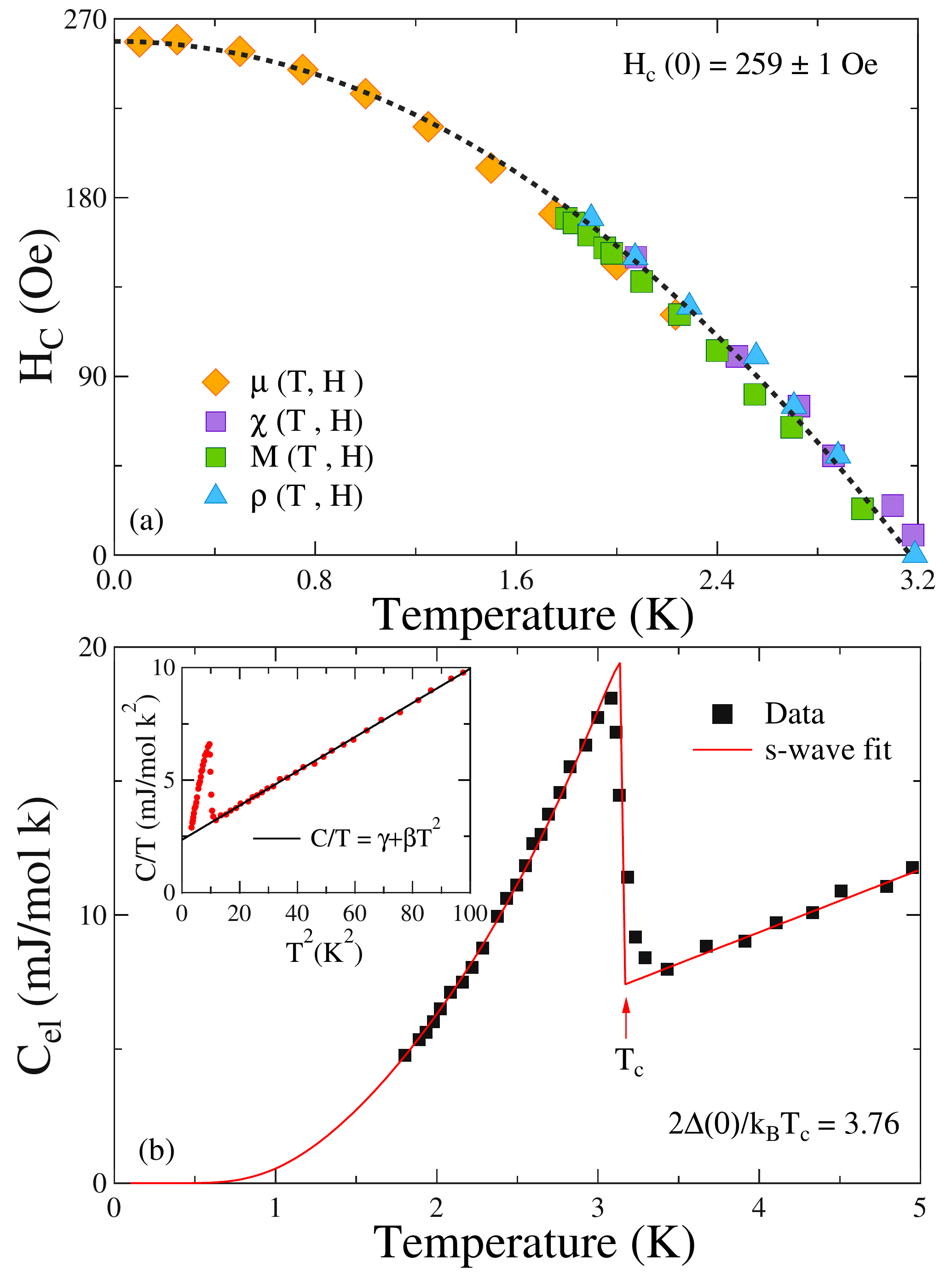}
\caption{\label{Fig5:Fig5}(a) Thermodynamic critical field $H_{c}$ as a function of temperature obtained from $\rho(T)$, M(H), $\chi(T)$ and muon experimental data. The solid line represents the fits using the function $H_{c}(T) = H_{c}(0)[1-(T/T_{c})^{2}]$. (b) Zero-field electronic specific heat data, $C_{el}$, as a function of temperature T. The solid red curve represents the single fully gapped superconductor for 2$\Delta$(0)/$k_{B}$ $T_{c}$ = 3.76. The inset shows the C/T vs $T^{2}$ data, where its fitted using the relation C/T = $\gamma$+$\beta$T$^{2}$.}
\end{figure} 
The magnetization measurement (M) as a function of field (H) at T = 1.8 K is shown in \figref{Fig4:Fig4}(a). The isothermal magnetization curve follows a near typical type-I or dirty type-II superconducting behavior. Decreasing the magnetic field from H = 300 Oe display an partial re-entrance of diamagntization as magnetic flux is expelled from the system. Such kind of magnetization data has been observed in other type-I superconducting materials such as: LaRhSi$_{3}$ \cite{VKA1}, LaMSi$_{3}$ (M= Pd,Pt)\cite{MSA3}, Ir$_{2}$Ga$_{9}$ \cite{KWS}. The absence of the typical step transition at critical field could be due to the demagnetization factor which often broadens the transition. Figure \ref{Fig4:Fig4}(b) displays the field dependence of the magnetization, M(H), performed at different temperatures from 1.8 K $\le$ T $\le$ 4 K to determine the $H-T$ phase diagram. The value for thermodynamic critical field $H_{c}$(T) is defined as the field at which the system goes from the superconducting state to the normal state. For example: $H_{c}$ = 168 Oe at T = 1.8 K [see \figref{Fig4:Fig4}(a)]. The resulting values of $H_{c}$ determined in this manner for different temperatures are summarized in \figref{Fig5:Fig5}(a) together with the values determined from the $\rho(T)$, $\chi(T)$ and muon data. The $H_{c}$(T) can be described by the conventional relation 
\begin{equation}
H_{c} = H_{c}(0)\left[1-\left(\frac{T}{T_{c}}\right)^{2}\right]
\label{eq1}
\end{equation}
where $T_{c}$ superconducting transition temperature and $H_{c}$(0) is critical field value for T = 0. Dotted black curve shows the best fit for the data which yields $H_{c}$(0) = 259 $\pm$ 1 Oe.\\ 
Figure \ref{Fig5:Fig5}(b) shows the specific heat data of AuBe measured in zero applied magnetic field. The sharp jump around $T_{c}$ $\approx$ 3.17 K confirms the intrinsic nature of superconductivity in this compound. The plot C/T versus $T^{2}$ is shown in the inset of \figref{Fig5:Fig5}(b). In the normal state, the data is best represented by $C/T = \gamma_{n}+\beta T^{2}$, where $\gamma_{n}$ is the Sommerfeld coefficient and the second term $\beta$ is the contribution from lattice. The solid black curve in the inset represents the fit to the data, yields $\gamma_{n}$ = 2.35 $\pm$ 0.02 mJ mol$^{-1}$ K$^{-2}$ and $\beta$ = 0.076 $\pm$ 0.003 mJ mol$^{-1}$ K$^{-4}$. The value for $\gamma_{n}$ was used to determine the density of states at the Fermi level $D_{c}(E_{\mathrm{F}})$ using the relation $\gamma_{n} =( \pi^{2}k_{B}^{2}D_{c}(E_{\mathrm{F}}))/3$, where $E_{\mathrm{F}}$ is the Fermi energy.  For  $\gamma_{n}$ = 2.35 $\pm$ 0.02 mJ mol$^{-1}$ K$^{-2}$, it yields $D_{c}(E_{\mathrm{F}})$ $\approx$ 1.0 $\pm$ 0.1 states eV$^{-1}$ f.u.$^{-1}$. The Debye temperature is given by $\theta_{D}$ = $\left(12\pi^{4}RN/5\beta\right)^{1/3}$, where using R = 8.314 J mol$^{-1}$ K$^{-1}$ and N = 2, yields $\theta_{D}$ = 370 $\pm$ 5 K. The value of $\theta_{D}$ = 370 K can be used to calculate the electron-phonon coupling constant $\lambda_{e-ph}$  using the McMillan formula \cite{WLM},
\begin{equation}
\lambda_{e-ph} = \frac{1.04+\mu^{*}\mathrm{ln}(\theta_{D}/1.45T_{c})}{(1-0.62\mu^{*})\mathrm{ln}(\theta_{D}/1.45T_{c})-1.04 } ,
\label{eqn4:ld}
\end{equation}                       
where $\mu^{*}$ represents the repulsive screened Coulomb potential, usually given by $\mu^{*}$ = 0.13. With $T_{c}$ = 3.17 K and $\theta_{D}$ = 370 K, we obtained $\lambda_{e-ph}$ $\simeq$ 0.54, which is similar to other weakly coupled NCS superconductors \cite{VKA1,MSA3,NBOS,ABK, SAS}. The bare-band effective mass m$^{*}_{band}$ can be related to m$^{*}$, which contains enhancements from the many-body electron phonon interactions m$^{*}$ = m$^{*}_{band}$(1+$\lambda_{e-ph}$) \cite{GG}. Using $\lambda_{e-ph}$ = 0.54 and assuming m$^{*}_{band}$ = m$_{e}$ yields m$^{*}$ = 1.54  m$_{e}$.\\  
The electronic contribution ($C_{el}$) to the specific heat determined by subtracting the phononic contribution from the measured specific heat data, i.e., 
$C_{el} = C-\beta T^{3}$, shown in the main panel of \figref{Fig5:Fig5}(b). The value for the specific heat jump is $\frac{\Delta C}{\gamma_{n}T_{c}}$ $\approx$ 1.51, which is close to the value for a BCS isotropic gap superconductor (= 1.43). This  indicates weakly-coupled superconductivity in AuBe, consistent with the $\lambda_{e-ph}$ value obtained above. The temperature dependence of the specific heat data in the superconducting state can best be described by the BCS expression for the normalized entropy S written as
\begin{equation}
\frac{S}{\gamma_{n}T_{c}} = -\frac{6}{\pi^2}\left(\frac{\Delta(0)}{k_{B}T_{c}}\right)\int_{0}^{\infty}[ \textit{f}\ln(f)+(1-f)\ln(1-f)]dy ,
\label{eqn5:s}
\end{equation}
where $\textit{f}$($\xi$) = [exp($\textit{E}$($\xi$)/$k_{B}T$)+1]$^{-1}$ is the Fermi function, $\textit{E}$($\xi$) = $\sqrt{\xi^{2}+\Delta^{2}(t)}$, where $\xi$ is the energy of normal electrons measured relative to the Fermi energy, $\textit{y}$ = $\xi/\Delta(0)$, $\mathit{t = T/T_{c}}$, and $\Delta(t)$ = tanh[1.82(1.018(($\mathit{1/t}$)-1))$^{0.51}$] is the BCS approximation for the temperature dependence of the energy gap. The normalized electronic specific heat is calculated by
\begin{equation}
\frac{C_{el}}{\gamma_{n}T_{c}} = t\frac{d(S/\gamma_{n}T_{c})}{dt}.
\label{eqn6:Cel}
\end{equation}
Fitting the low temperature specific heat data using this model as shown by the solid red line in \figref{Fig5:Fig5}(b), yields $\Delta(0)/k_{B}T_{c} $ = 1.82 $\pm$ 0.2. This is consistent with the value for a BCS superconductor $\alpha_{BCS}$ = 1.764 in the weak coupling limit. Therefore, good agreement between the measured data (black symbols) and the BCS fit (solid red line), confirms an isotropic fully gapped BCS superconductivity in AuBe.\\   
\begin{figure}
\includegraphics[width=1.0\columnwidth]{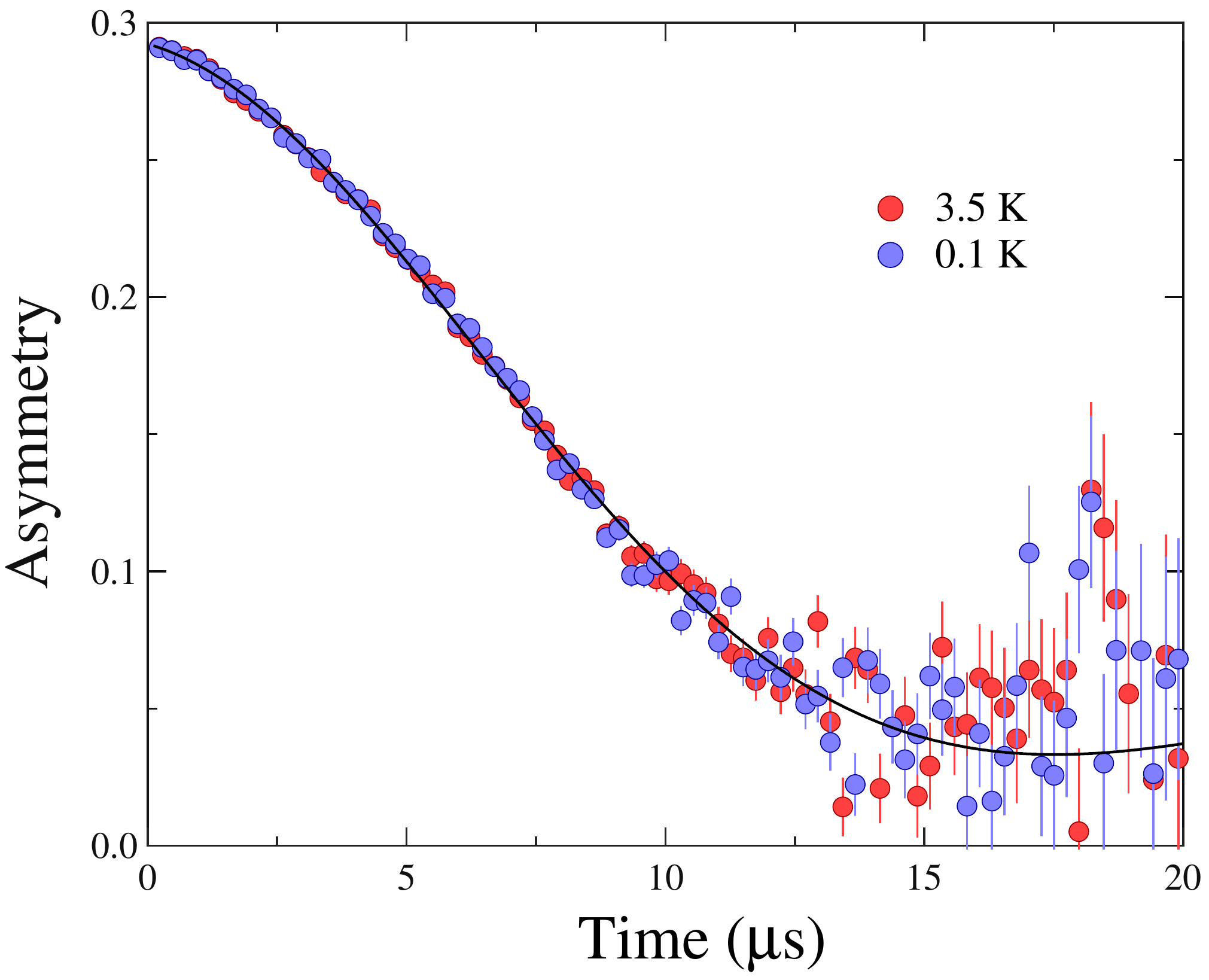}
\caption{\label{Fig6:Fig6} Zero-field $\mu$SR spectra collected below (0.1 K) and above (3.5 K) the superconducting transition temperature. The solid lines are the fits to Guassian Kubo-Toyabe (KT) function given in Eq. \ref{eqn3}.}
\end{figure}
The time evolution of muon spin relaxation in zero field is shown in \figref{Fig6:Fig6} for temperatures both above 3.5 K ($>$ $T_{c}$) and below 0.1 K ($<$ $T_{c}$) the transition temperature $T_{c}$. The ZF-$\mu$SR spectra are well fitted with the following function:
\begin{equation}
G(t)= A_{1}\mathrm{exp}(-\lambda t)G_{\mathrm{KT}}(t)+A_{\mathrm{BG}} ,
\label{eqn3}
\end{equation}
where $A_{1}$ is the initial asymmetry, $\lambda$ is the electronic relaxation rate and $A_{\mathrm{BG}}$ is the time-independent background contribution. The G$_{\mathrm{KT}}$(t) function is the Gaussian Kubo-Toyabe function given by
\begin{eqnarray}
G_{\mathrm{KT}}(t) &=&\frac{1}{3}+\frac{2}{3}(1-\sigma^{2}_{\mathrm{ZF}}t^{2})\mathrm{exp}\left(-\frac{\sigma^{2}_{\mathrm{ZF}}t^{2}}{2}\right),
\label{eqn4}
\end{eqnarray}
where $\sigma_{ZF}/\gamma_{\mu}$ is the local field distribution width, $\gamma_{\mu}$ = 13.553 MHz/T being the muon gyromagnetic ratio. In systems where the superconducting state breaks time-reversal symmetry, spontaneous magnetic moments arise below $T_{c}$ and an increase may be observed in either $\sigma_{ZF}$ or $\lambda$. It is evident from the ZF-$\mu$SR spectra that there is no noticeable change in the relaxation rates at either side of superconducting transition. This indicates that the time-reversal symmetry is preserved in the SC phase within the experimental accuracy.\\
\begin{figure}
\includegraphics[width=1.0\columnwidth]{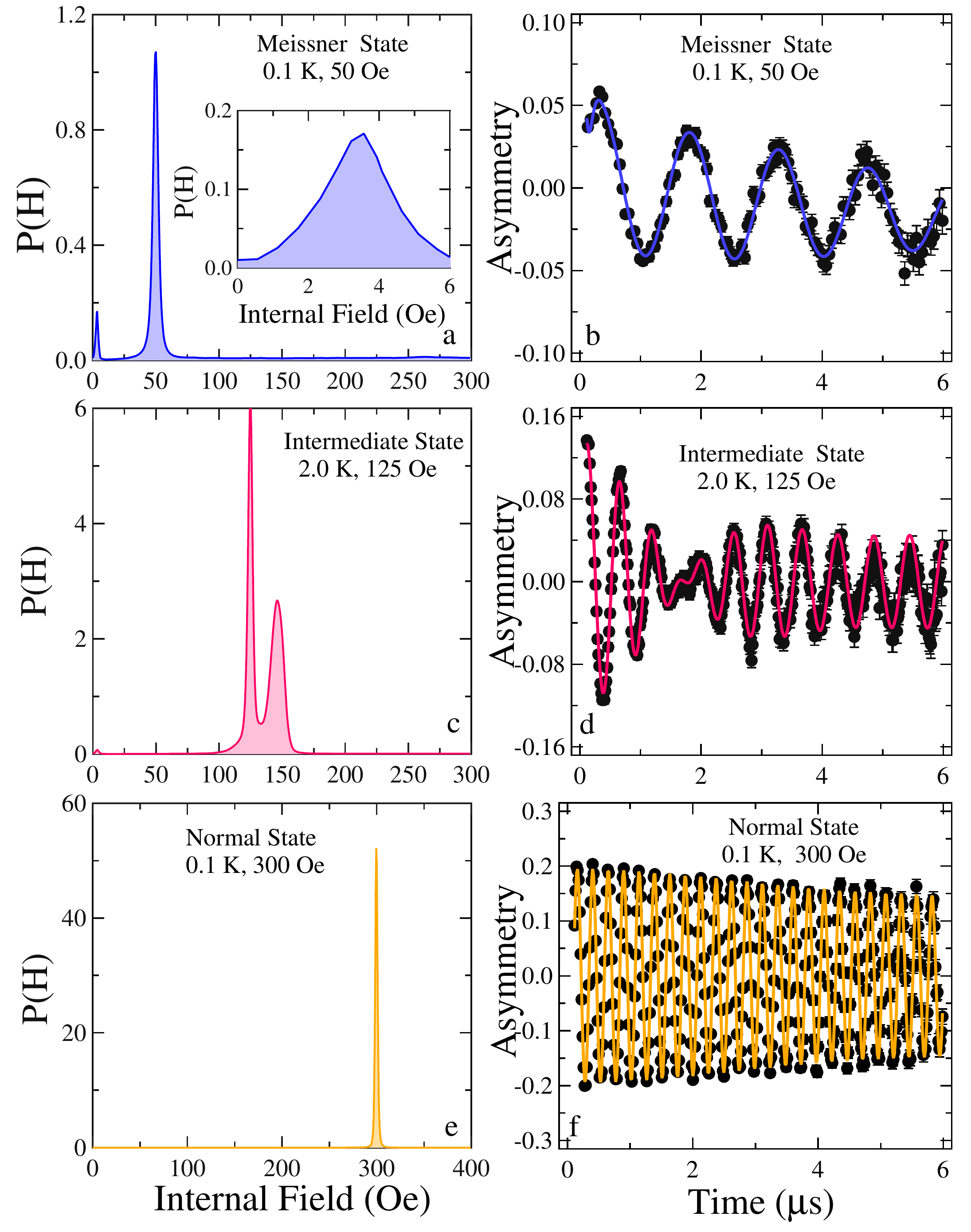}
\caption{\label{Fig7:Fig7} Field distributions and TF-$\mu$SR signals. Field distribution of the local field probed by the muons, P(H), obtained by MaxEnt transformation of the TF-$\mu$SR time spectra at different temperatures and applied field. The figure illustrates typical signal observed in the (a) Meissner, (c) Intermediate state and (e) Normal state. b, d, and f show the TF-$\mu$SR time spectra for the corresponding states. The solid lines are fits to the data using a fit function, described in Eq. 4}
\end{figure} 
 The transverse-field $\mu$SR data were collected after cooling the sample in an applied field from the normal state into the superconducting state. The TF-$\mu$SR precession signals were obtained in different applied fields up to 300 Oe in several temperatures above and below the transition temperature. Figure \ref{Fig7:Fig7} show the typical MaxEnt results of the magnetic field distribution, extracted from the TF-$\mu$SR time spectra \cite{MAX}, in the (a) Meissner, (c) Intermediate and (e) Normal state. Figure \ref{Fig7:Fig7} (b), (d), and (f) show the TF-$\mu$SR time spectra for the corresponding states. At H = 50 Oe and T = 0.1 K, AuBe is in the Meissner state. This state is well reflected in the MaxEnt results where we observe a peak near $H_{int}$ $\approx$ 3 Oe, which corresponds to the Au nuclear moments [see inset \figref{Fig7:Fig7}(a)]. In the main panel of \figref{Fig7:Fig7}(a), the second peak around H = 50 Oe is a background signal mainly due to the muons stopping in the other parts of the sample holder. The absence of any additional peak implies that the magnetic field is completely expelled from the body of the superconductor. Figure \ref{Fig7:Fig7}(b) shows the TF-$\mu$SR spectra in the Meissner state (H = 50 Oe, T = 0.1 K) where the weak decay suggests the Kubo-Toyabe behaviour associated with the nuclear moments. Interestingly, there is a considerable reduction in the intial asymmetry. The loss of initial asymmetry as observed in our TF-$\mu$SR spectra of AuBe is similar to that observed in LaRhSi$_{3}$ \cite{VKA1}, LaNiSn \cite{LNS}, which exhibits type-I superconductivity. At the higher applied field or temperature near $T_{c}$ the initial asymmetry recovers to its full maximum value.\\% The spectra in the \figref{Fig7:Fig7}(b) were fitted using the combination of Eqs. \eqref{eqn3} and \eqref{eqn5}. The maximum entropy spectra for the 50 Oe data at 0.1 K is showin in \figref{Fig7:Fig7}(a) where there is a peak in P(H) near H = 0 Oe, as expected for a sample in Meissner state. The absence of any additional peak implies that the magnetic field is completely excluded from the sample.\\ 
The TF-$\mu$SR spectra at H = 125 Oe field and T = 2.0 K shown in \figref{Fig7:Fig7}(d). We analyze the spectra using the following function:
\begin{eqnarray}
G_{\mathrm{z}}(t)=  G(t)+\sum_{i=1}^N A_{i}\exp\left(-\frac{1}{2}\sigma_i^2t^2\right)\cos(\gamma_\mu B_it+\phi),
\label{eqn5}
\end{eqnarray}
where G(t) is the Eqn. \eqref{eqn3}, $A_{i}$ is the initial asymmetry, $\sigma_i$ is the Gaussian relaxation rate, $\gamma_{\mu}/2\pi$ = 135.5 MHz/T is the muon gyromagnetic ratio , common phase offset $\phi$, and $B_i$ is the first moment for the $i$th component of the field distribution. We found that the asymmetry spectra can best be described by two oscillating functions (N=2). In these fits, the i = 1 depolarization component was fixed to $\sigma_{1}$ = 0, which corresponds to a background term arising from those muons stopping in the silver sample holder as they do not appreciably depolarize over the time-scale of the experiment. From these fits, we obtain the value of the internal magnetic fields 125 Oe and 146 Oe. The former value of the field is the same as the applied field (from the silver holder), while the latter value can be taken as an estimate of the critical field ($H_{c}$) coming from the intermediate state of a type-I superconductor. Intermediate state of a type-I superconductor is induced by the non-zero demagnetization effects. In such situation, a stable coexistence of the flux-free regions and the regions of internal field $\approx$ $H_{c}$ arises even if the magnetic field applied, $H_{app}$, is considerably less than the critical field $H_{c}$. Muons implanted in these normal regions of the intermediate state will precess at a frequency corresponding to the field at the muon site which must be at least equal to $H_{c}$. Muons implanted in regions where magnetic flux is expelled will only be affected by nuclear moments. In the MaxEnt data shown in \figref{Fig7:Fig7}(c), two sharp peaks demonstrates the two field components. For a type-II superconductor in the mixed state, we expect a field component at a lower value than the applied field due to the establishment of the flux-line lattice (FLL). This is clearly absent for our sample. On the contrary, we observe that AuBe entering the intermediate state where the regions which are normal yields a peak near $H_{c}$ = 146 Oe in the MaxEnt data. This is a strong evidence for bulk type-I superconductivity in the compound. In the magnetic field of H = 300 Oe and T = 0.1 K, AuBe returns to the normal state. Here, the field penetrates the bulk of the sample completely, and we see a almost homogeneous field distribution in the TF-$\mu$SR time spectra as displayed in \figref{Fig7:Fig7}(f). Accordingly, a single sharp peak is observed at H = 300 Oe in the maximum entropy spectra in \figref{Fig7:Fig7}(e).\\
Figure \ref{Fig8:Fig8}(a) shows the internal field distribution, P(H), obtained by MaxEnt transformation at different applied fields between 50 Oe $\le$ H $\le$ 250 Oe at constant temperature T = 0.1 K. The internal field distribution at T = 0.1 K, clearly demonstrate the change from the Meissner state (H = 50 Oe) to the intermediate state (H = 150 Oe) to the normal state (H = 250 Oe) with increasing field. At the intermediate state for $H_{app}$ $>$ 50 Oe, P(H) shows an additional peak at $H_{int}$ $>$ $H_{app}$ corresponding to $H_{c}$. There is not a significant change in the $H_{c}$ peak position with the increase in applied magnetic field. It is noteworthy here that for $H_{app}$ $\ge$ 50 Oe we observed systematic decrease in P(H) near $H_{int}$ $\sim$ 3 Oe [see inset \figref{Fig8:Fig8}(a)]. This can be understood from the fact that at the low-field (50 Oe) system is in complete Meissner state, depicted by the topmost peak in the inset graph. As the field is increased system goes to the intermediate state, which consequently decreases the Meisnner volume. Since the P(H) peak near the low-field region arises from the Meisnner state, it is understandable that it decreases with the increase in applied field. At $H$ = 250 Oe the system goes to the normal state which is apparent from the presence of only one peak in the P(H) graph.\\ 
\begin{figure}
\includegraphics[width=1.0\columnwidth]{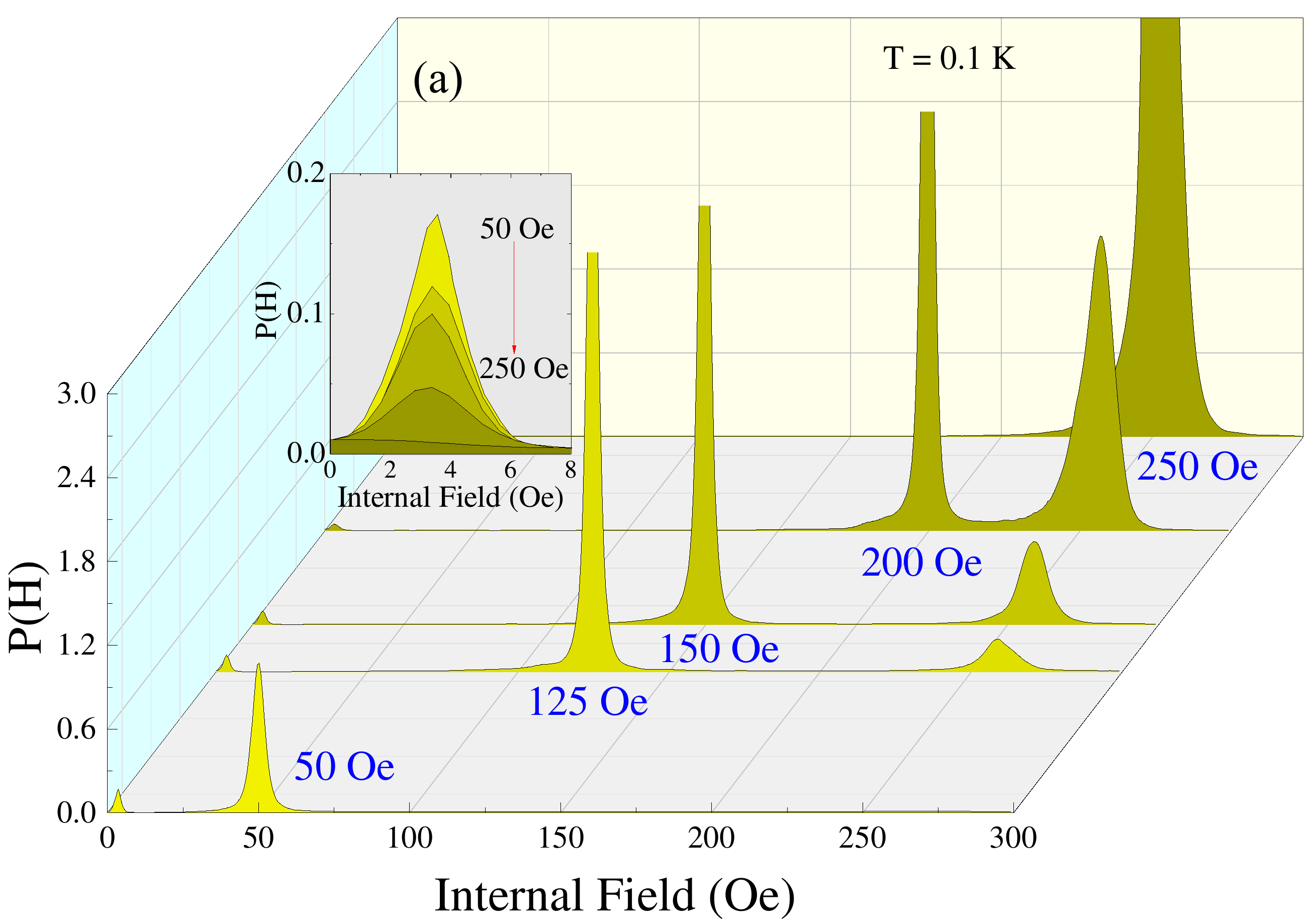}
\includegraphics[width=1.0\columnwidth]{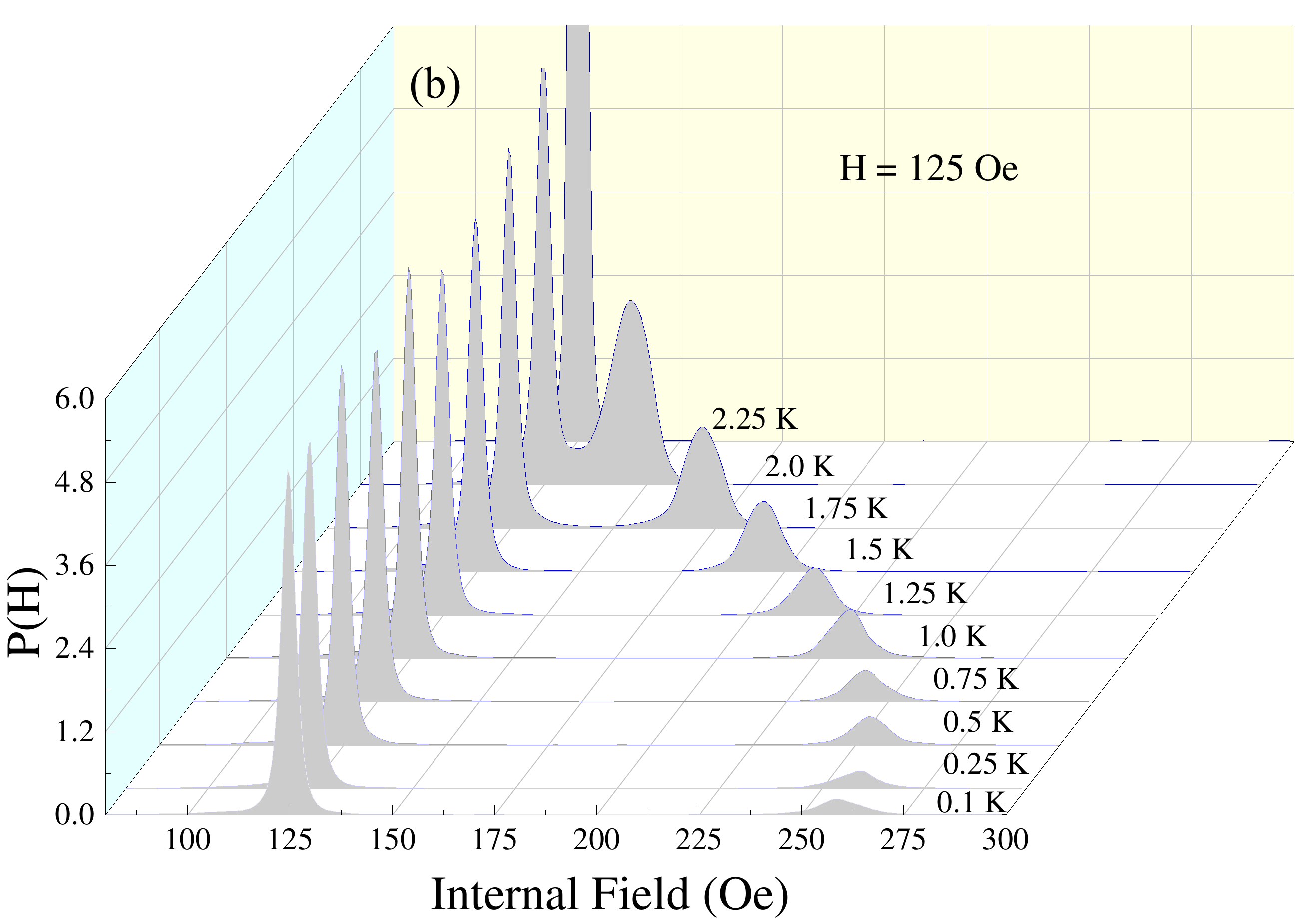}
\caption{\label{Fig8:Fig8}(a) Field distribution of the internal field probed by the muons, P(H), obtained by MaxEnt transformation at different fields between 50 Oe $\le$ H $\le$ 250 Oe at T = 0.1 K and (b) between 0.1 K $\le$ T $\le$ 2.25 K at applied field H = 125 Oe.}
\end{figure}
The maximum entropy spectra for TF-$\mu$SR in a range of temperatures between 0.1 K $\le$ T $\le$ 2.25 K are shown in \figref{Fig8:Fig8}(b). The sample was cooled in an applied field of H = 125 Oe, from above $T_{c}$ and the measurements were made while warming. At T = 0.1 K two different peaks appears in the P(H) spectra at $H_{app}$ = 125 Oe and and $H_{c}$ = 256 Oe, reminiscent of the intermediate state in a type-I superconductor. A peak also appears at the low-field region due to the Meissner state. Interestingly, as the temperature is increased the peak position for $H_{c}$ moves closer to $H_{app}$ with a subsequent increase in the magnitude of internal field. The observed behaviour is due to the increase in the volume of normal region in the intermediate state as the temperature is increased. Thus, we can deduce the $H$-$T$ phase diagram from the peak position of the critical field $H_{c}$. The obtained phase diagram is shown in \figref{Fig5:Fig5}(a) and are in good agreement with those obtained from the other measurement techniques. The critical field $H_{c}$(0) is 259 $\pm$ 1 Oe obtained after fitting Eqn. \eqref{eq1}.\\
We now investigate the superconducting parameters of AuBe. The London penetration depth is given by the relation $\lambda_{L}$ = (m$^{*}$/$\mu_{0}$ne$^{2}$)$^{1/2}$. Putting in the values m$^{*}$ = 2.8m$_{e}$ and n = 3.93 $\times$ 10$^{28}$ $m^{-3}$, yields $\lambda_{L}$ $\approx$ 450 $\text{\AA}$. The BCS coherence length can be evaluated using the expression $\xi_{0}$ = $\left(\frac{0.18\hbar v_{F}}{k_{B}T_{c}}\right)$ $\approx$ 1883 $\text{\AA}$. It can  be pointed out that $\xi_{0}$ is larger than the calculated mean free path $l$ (808.2 $\text{\AA}$), $l/\xi_{0}$ $\approx$ 0.43 $<<$ 1, indicating that the superconductivity in AuBe is in the dirty limit. In the dirty limit, the Ginzburg-Landau (GL) parameter $\kappa_{GL}$ = 0.715 $\lambda_{L}(0)/l$ \cite{MTIN}, which gives $\kappa_{GL}$ $\approx$ 0.4. $\kappa_{GL}$ is smaller than the value 1/$\sqrt{2}$ $\approx$ 0.707 separating type-I and type-II superconductivity, suggesting that AuBe is an type-I superconductor. Further using the relation $\kappa_{GL}$ = 7.49 $\times$ 10$^{3}$ $\rho_{0}$$\sqrt{\gamma_{nV}}$, with $\rho_{0}$ in $\Omega$ cm and $\gamma_{nV}$ in units erg/cm$^{3}$K$^{2}$, we obtained $\kappa_{GL}$ = 0.4, consistent with the value obtained above. The effective magnetic penetration depth $\lambda_{\mathrm{eff}}$ is equal to 821 \text{\AA}, which is calculated using the relation $\lambda_{\mathrm{eff}}$ = $\lambda_{L}$$\sqrt{1+\xi_{0}/l}$. In addition, the Ginzburg-Landau coherence length $\xi$(0) determined from the relation $\kappa_{GL}$ = $\lambda_{\mathrm{eff}}$/$\xi$(0), which yields $\xi$(0) = 2052 for $\kappa_{GL}$ = 0.4.\\
\begin{table}[h!]
\caption{Normal and superconducting properties of noncentrosymmetric superconductor AuBe}
\label{elec propr}
\begin{center}
\begin{tabular*}{1.0\columnwidth}{l@{\extracolsep{\fill}}lll}\hline\hline
Parameter& unit& value\\
\hline
\\[0.5ex]                                  
$T_{c}$& K & 3.2 \\
$\gamma_{n}$& mJ mol$^{-1}$ K$^{-1}$ &2.35\\
$\theta_{D}$& K &370\\
$\Delta C_{el}/\gamma_{n}T_{c}$& & 1.51\\
$H_{c}$& Oe& 259\\
$\kappa_{GL}$& & 0.4\\
$\xi$(0)& \text{\AA} & 2052\\
$\xi_{0}$& \text{\AA} & 1883\\
$l$& \text{\AA}& 808.2\\
$\lambda_{L}$& \text{\AA} & 450 \\
$\lambda_{\mathrm{eff}}$& \text{\AA} & 821 \\
$T_{\mathrm{F}}$& K& 17450\\
$T_{c}/T_{\mathrm{F}}$& & 0.00018\\
\\[0.5ex]
\hline\hline
\end{tabular*}
\par\medskip\footnotesize
\end{center}
\end{table} 
According to Uemura et al. \cite{YJU1,YJU2,YJU3} superconductors can be conveniently classified according to their $\frac{T_{c}}{T_{F}}$ ratio. It was shown that for the unconventional superconductors such as heavy-fermion, high-$T_{c}$, organic superconductors, and iron-based superconductors this ratio falls in the range 0.01 $\leq$ $\frac{T_{c}}{T_{F}}$ $\leq$ 0.1. In \figref{Fig10:Fig10}, the region between the green solid lines represents the band of unconventional superconductors. For a 3D system the Fermi temperature T$_{F}$ is given by the relation
\begin{equation}
 k_{B}T_{F} = \frac{\hbar^{2}k_{F}^{2}}{2m^{*}}, 
\label{eqn13:tf}
\end{equation}
where $k_{F}$ is the Fermi vector. Using the estimated value of $k_{F}$ for AuBe in Eq. \ref{eqn13:tf}, it yields $T_{F}$ = 17450 K, giving $\frac{T_{c}}{T_{F}}$ $\approx$ 0.00018. AuBe is located well outside the range of unconventional superconductors and close to the vicinity of elemental superconductors which are type-I BCS superconductors as shown by a solid red marker in \figref{Fig10:Fig10}. This potentially suggesting conventional mechanism of superconductivity in AuBe.
\begin{figure}
\includegraphics[width=1.0\columnwidth]{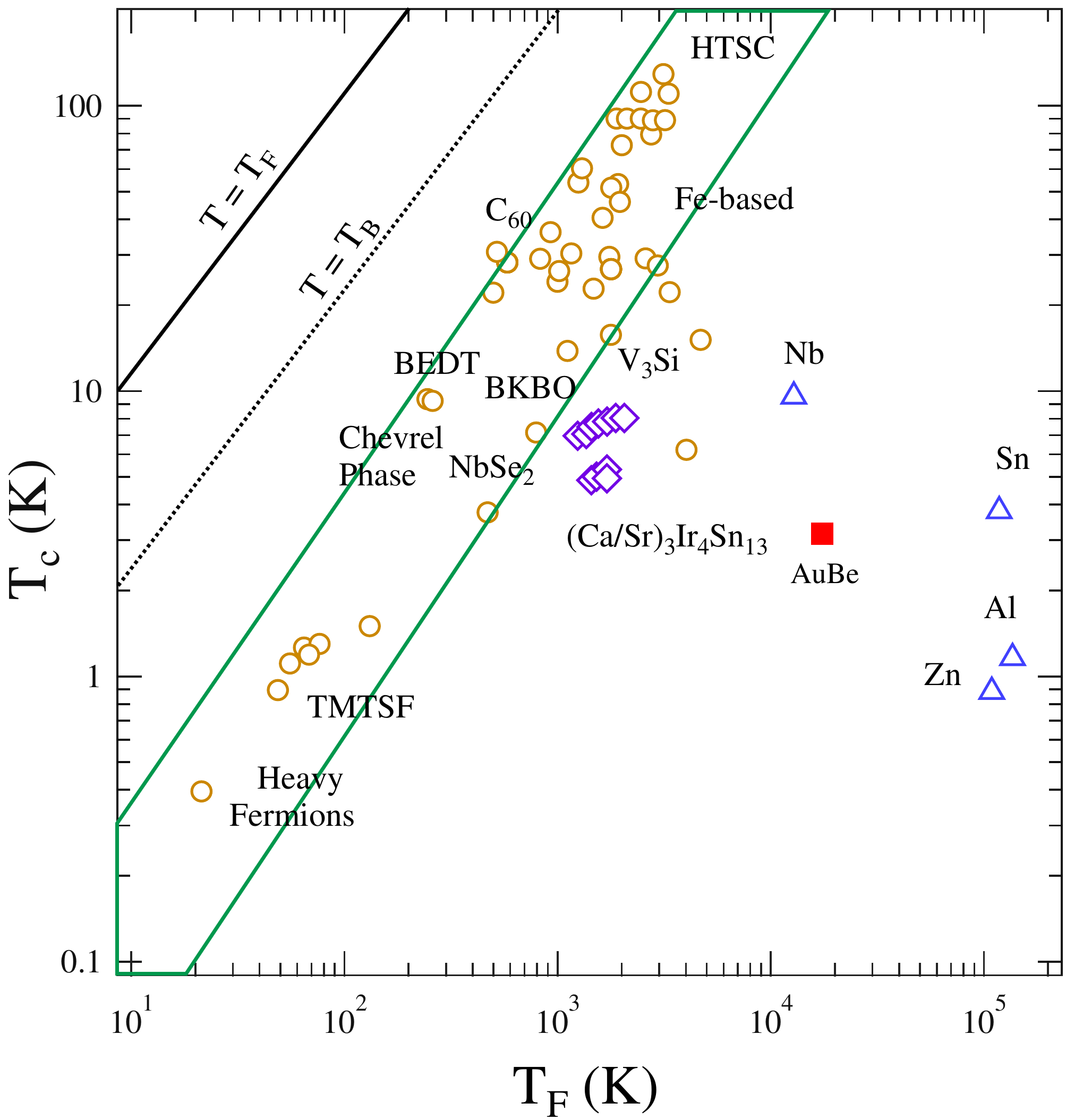}
\caption{\label{Fig10:Fig10} The Uemura plot showing the superconducting transition temperature $T_{c}$ vs the Fermi temperature $T_{F}$, where AuBe is shown as a solid red square marker well outside the range of band of unconventional superconductors. The region displayed by the soild green lines represents the band of unconventionality were obtained from Ref. \cite{YJU1,YJU2,YJU3}.}
\end{figure}

\section{Conclusions}

To summarize, we have investigated the superconducting properties of cubic noncentrosymmetric superconductor AuBe using magnetization, resistivity, specific heat and $\mu$SR measurements. AuBe goes to superconducting transition at $T_{c}$ = 3.2 K. The specific heat data measured in zero applied field shows BCS superconductivity in the weak-coupling regime. Interestingly, magnetization measurements along with the calculated superconducting parameters suggest a type-I superconductivity in this compound, in contrast to the earlier reports which showed type-II superconductivity. The thermodynamic critical field $H_{c}$ $\approx$ 259 Oe and the GL parameter $\kappa_{GL}$ is 0.4 $<<$ 1/$\sqrt{2}$, again confirming type-I superconductivity in this system. The microscopic study of superconductivity in AuBe were done $\mu$SR measurements. The TF-$\mu$SR data was transformed into the probability of field versus field graph, P(H), using the MaxEnt algorithm. The field components obtained doesn't show any signature of mixed state to suggest type-II superconductivity in AuBe. Notably, in a type-II superconductor, the peak around the applied field broadens and an additional shoulder in the distribution is observed at lower fields due to FLL in the mixed state. However, this feature was not observed for any temperature or field. In fact, an internal field at a greater frequency than the applied field is observed which is strong evidence for bulk type-I superconductivity in the compound. Furthermore, the ZF-$\mu$SR analysis indicates that the time-reversal symmetry is preserved in the superconducting state. 
\section{Acknowledgments}
R.~P.~S.\ acknowledges Science and Engineering Research Board, Government of India for the Ramanujan Fellowship through Grant No. SR/S2/RJN-832012. We thank Newton Bhabha funding and ISIS, STFC, UK, for the muon beam time to conduct the $\mu$SR experiments [DOI: 10.5286/ISIS.E.67770571].

\end{document}